\documentstyle[11pt,epsfig,amstex,amssymb]{article}                                           
\newcommand{\la}{$\Lambda~$}                                         
\newcommand{\A}{$\alpha_S~$}                                         
\newcommand{\q}{$Q^2~$}                                            
\newcommand{\Q}{$Q^2_0~$}                                            
                                            
\newcommand{\f}{$F_2~$}                                            
\newcommand{\F}{$F_2(x,Q^2)~$}                                            
\newcommand{\beq}{\begin{equation}}                                            
\newcommand{\eeq}{\end{equation}}                                             
\begin{document}                                                               
\begin{titlepage}                                                              
\hfill
\hspace*{\fill}
\begin{flushleft}      
       
\noindent     
{\tt DESY 96-063    \hfill    ISSN 0418-9833} \\    
{\tt April 1996} \\          
\end{flushleft}         
\vspace*{2.cm}                                                 
                
\begin{center}                                                                 
\begin{LARGE}                                                                  
{\bf  On the Asymptotic Behaviour
of  {\boldmath $F_2(x,Q^2)$ } }\\
\end{LARGE}                                                                    
\vspace{3cm}  
\begin{Large}                                                                 
{
A.DeRoeck$^1$, M.Klein$^2$ and Th.Naumann$^2$ 
}\\ 
\vspace{1cm}                                                   
$^1$ DESY, Notkestr.85, D-22607 Hamburg \\
$^2$ DESY Zeuthen, Platanenallee 6, D-15738 Zeuthen                           
\\
\end{Large}
\end{center}                                                                   
\vspace*{5.cm}                          
\begin{center}                                       
{\bf Abstract}
\end{center}                                                                
\begin{quotation}                                                              
We discuss how the proton structure function \F
is described in the HERA kinematic range by
double asymptotic expressions for low $x$ and large \q. 
From a NLO double asymptotic approximation of recent data 
from the H1 experiment at HERA
we extract the strong coupling constant
$\alpha_S(M^2_Z)=0.113\pm0.002(stat)\pm0.007(syst)$.
The additional theoretical error can be as large as 0.007.
\end{quotation}                                                                
\vfill                                                                         
\vfill
\noindent
\vspace{1cm}
\end{titlepage}                                                                
\newpage                                                                       
\vspace{2cm}                                                                   
\noindent                                                                      

Previous fixed target lepton-nucleon scattering experiments \cite{NMC,BCDMS}
measured the proton structure function \F
at Bjorken-$x \gtrsim 0.01$ and
were  sensitive to the valence quark content of the nucleon. 
HERA experiments collect data at values of $x$ low enough to
neglect valence quarks and at values of \q large enough to
apply perturbation theory.  Here
\q is the four-momentum transfer squared from the incoming lepton to
the proton.
The proton structure function \f strongly rises with decreasing
$x$ over the whole kinematic range
accessible to the HERA experiments H1 and ZEUS \cite{H1,ZEUS}.
Such a rise was predicted more than twenty years ago~\cite{Ruju} 
from the leading order renormalization
group equations of perturbative QCD. 
Ball and Forte pursued these ideas~\cite{Ball}
and proposed a way to demonstrate
that the low $x$ data at HERA exhibit scaling properties 
dominantly generated by  QCD radiation. They gave an expression for
\F in the double asymptotic limit of low $x$ and large \q. 
The recent \F measurement of  H1 is in accord with such 
scaling behaviour. Hence these data are expected to be sensitive to
the fundamental  QCD evolution dynamics, and not to depend on unknown
starting distributions. The formalism of the double asymptotic
analysis was extended to two loops in~\cite{Ball2}.


We define the logarithmic QCD evolution variables
\[ t = \ln (Q^2  /\Lambda^2) \hspace{10mm} \mbox{and} \hspace{10mm} 
 \xi = \ln (1/x) \]
where $\Lambda$ is the QCD mass scale,
and the evolution length $T$ of $\alpha_S(Q^2)$ from
a starting point $Q^2_0$ to $Q^2$
\[ T = \ln\;(\alpha_S(Q_0^2)/\alpha_S(Q^2)). \]
To leading order $T$ is simply 
 $ T = 4\pi/b_0 \int_{t_0}^{t} dt\;\alpha_S(t) = \ln (t/t_0). $
The $T$ dependence of \f allows to determine $\alpha_S$.
Here $b_0= 11-2n_f/3$ is the leading coefficient of the renormalization
group equations, with $n_f$ the number of flavours.


For large $t$ and $\xi$ and an \F which at \Q is not too
singular in $x$ the NLO double asymptotic expression for \F is \cite{Ball2}
\begin{equation}
F_2 \sim N_F~(1-f_{NLO})~\exp\;(2\gamma \sqrt{\xi T} -\delta T + \frac{1}{4} \ln T - \frac{3}{4} \ln \xi).
\end{equation}
with
\[ \gamma^2 = 12/b_0 \hspace{3cm}  \delta = (11+2n_f/27)/b_0 . \]
The normalization coefficient is $N_F = \sqrt{\gamma/\pi} ~5n_f/324 .$
For $n_f=4$ this gives $\gamma=6/5$ , $\delta=4/3$ and $N_F=0.038$.
The NLO correction term $f_{NLO}$ is
\[f_{NLO}=\frac{\sqrt{\xi/T}}{2\pi\gamma}[\epsilon(\alpha_S(Q^2_0)-\alpha_S(Q^2))-13\alpha_S(Q^2)] \]
with $\epsilon = (206n_f/27+6b_1/b_0)/b_0$ and 
$  b_1 = 102-38n_f/3 $.
%
The leading order formula is recovered 
setting $f_{NLO}=0$ and $b_1=0$.
Defining the leading exponent as
$\lambda \xi = 2\gamma \sqrt{\xi T}$
and the subleading term as
$\alpha = -\delta T + 1/4 \ln T - 3/4 \ln \xi $
we rewrite \f as
\begin{equation}
F_2 \sim x^{-\lambda} e^{\alpha}.
\end{equation}
The leading term in the double asymptotic formula for \f corresponds to 
the double leading log approximation DLL \cite{GLR} of the
DGLAP equations \cite{DGLAP}. It generates
the growth  of the structure function 
with falling $x$ proportional to $x^{-\lambda}$.
The subleading term falls with $x$ but slower than the leading
term grows.

Expression (1) for the structure function is a prediction
of QCD in the limit of small $x$ and large \q.
It does not depend on the shape of the parton distribution functions
at \Q if these are sufficiently soft.
The starting point \Q of the QCD evolution is a free parameter.
Conventional QCD analyses of structure functions 
\cite{QCDH1,QCDZEUS,MRS,CTEQ}
fit  input parton distributions at some \Q
and evolve them to larger \q assuming a singular behaviour $F_2 \sim
x^{-\lambda}$ of the input distributions.
On the contrary, the approach of
GRV \cite{GRV} starts with non-singular, valence-like input distributions
at some low \Q and generates the steep rise of the
\f from QCD dynamics.

Independently of a set of input parton distributions 
we now determine \la and \A 
fitting expression (1) for \F to
the latest measurement of the proton structure function by the
H1 experiment \cite{H1} at HERA.
When passing the heavy flavour thresholds we adjust $\Lambda$
according to \cite{marc}.
In order to reach the asymptotic region in $T$ and $\xi$ we
demand $Q^2 \geq Q^2_{min} = 5 $ GeV$^2 > Q^2_0$ and $x<0.1$
which corresponds to  $\xi T \gtrsim 2$ approximately.
The fit using the total errors gives the central value $\Lambda= 248$
MeV with a $\chi ^2 / ndf = 122/145$ and $Q^2_0 = 1.12 $ GeV$^2$.
The $\Lambda$ symbol is taken as an abbreviation for 
$\Lambda_{\overline{MS}}^{(4)}$\cite{marc,bern2}.
The result is shown as a full line in Fig. 1 and describes the
data for $Q^2\ge 5 $ GeV$^2$ reasonably well.
A fit of the LO expression of equation (1) to the 
\f data is found to drop
too fast with $x$. The $\chi^2/ndf$ increases by a factor of two.

The statistical error is $\delta\Lambda_{stat}= 23$ MeV
from a fit with the statistical errors only.
Using the full systematic error matrix of H1 we then vary each of the
measured quantities entering the \f analysis.
From the variations of the \la determination with the varied \f
we get the full systematic error $\delta\Lambda_{syst}=83$ MeV .
In order to estimate the systematic error coming from the kinematic cuts
we also varied $Q^2_{min}$ from 5 to 60 GeV$^2$ and asked for $x<0.05$
or $x>0.001$. We also replaced the upper $x$ limit by demanding that $\xi T > 2$.
The resulting \la variation is 46 MeV.
The \la determination translates into
$\alpha_S(M^2_Z)=0.113\pm0.002(stat)\pm0.007(syst)$.

The heavy flavour treatment causes a theoretical uncertainty. 
Since we work at $Q^2$ above the charm threshold only 
the bottom quark mass $m_b$ enters the
uncertainty. We chose an effective $m_b=4.74$ GeV according to
\cite{pdg}. Taking $m_b=4.50$ GeV instead changes \la by 13 MeV.
Using a different prescription~\cite {bern} for the transition between
$\Lambda^{(n_f)}$ and $\Lambda^{(n_f-1)}$ changes \la by 7 MeV.
Changing as an extreme assumption the mass scale from \q to $W^2$ 
changes \la by 15 MeV.
The heavy flavour treatment thus yields a systematic error on \A of 0.002.
In order to check the influence of the asymptotic expression (1)
on the \A determination
we also fitted it to the H1 QCD fit 
\cite{H1} and to a set of GRV parametrizations with varying \la \cite{vogt} for $x>0.001$.
For the \Q obtained above the \la values from these fits
follow the variations of the values used in the QCD evolutions of
these parton densities and differ by at most 34 MeV from the input
values. This corresponds to an \A error of 0.003. 
The largest theoretical error, however, is 
anticipated to come from higher order
perturbative corrections and subleading terms.
This error is quoted to be in the range of 0.004 \cite{scheme} to
0.007 \cite{alpha}. 
We conservatively quote our result with a theoretical error of 0.007.

We finally determine a QCD inspired parametrization of $F_2$
which is valid down to the lowest $Q^2$ values.
Since the NLO fit shows that the structure function has not evolved
steep enough  below $Q^2 = 5$ GeV$^2$  we
attempted to fit also the constant of the leading exponent 
in equation (1)
fixing $F_2 = ax^{-\lambda}$ with $\lambda = c \sqrt{T/\xi}$.
Including now the data at $1.5<Q^2<5$ GeV$^2$
we get values for $a$ and $c$ very close to $N_f$ and 
$\gamma$ for four flavours.
Hence we fix $a=N_f$ and $c=\gamma$ and fit $F_2 = N_fx^{-\gamma \sqrt{T/\xi}}$
with $n_f = 4$ for $x < 0.1$. 
The result is shown in Fig. 1 as a dashed line. The values of the
only two fit parameters are $Q^2_0 = 0.365\pm 0.026(stat)\pm 0.048(syst)$
and $\Lambda = (243\pm 13\pm 23$) MeV. 
The two parameters resemble the GRV \Q scale and the QCD parameter \la.
The $\chi^2/ndf$ of the fit using full errors is 109/167.
Leaving the flavour number $n_f$ free gives very similar \Q and 
$\Lambda$ values, yielding $n_f = 3.83 \pm 0.18\pm0.34$ with almost the
same $\chi^2/ndf=108/166$. 
For $x<0.1$ this fit describes the \f data surprisingly well with the two
parameters \Q and $\Lambda$ only.



\section*{Summary}
For $Q^2 \geq 5$ GeV$^2$ 
a NLO double asymptotic expression for \F describes the HERA
data well. We use its \q dependence to extract 
$\alpha_S(M^2_Z)=0.113\pm0.002(stat)\pm0.007(syst)$.
We quote a theory error of 0.007 which is subject to uncertainties.
We found a simple QCD inspired parametrization 
$F_2 = N_fe^{-\gamma\sqrt{T/\xi}}$ for the proton structure function
which is valid over more than three orders of magnitude in $x$ and \q
including the region $Q^2<5$ GeV$^2$ and depends on two fitted scales
\Q and $\Lambda$ only.

\section*{Acknowledgement}
We thank R. Ball, J.Bl\"umlein, S.Forte and A. Vogt for helpful discussions.

\newpage
\begin{figure}[htbp] 
\begin{center}                                                                 
\begin{picture}(160,160)
\put(-15,-50)
{\epsfig{file=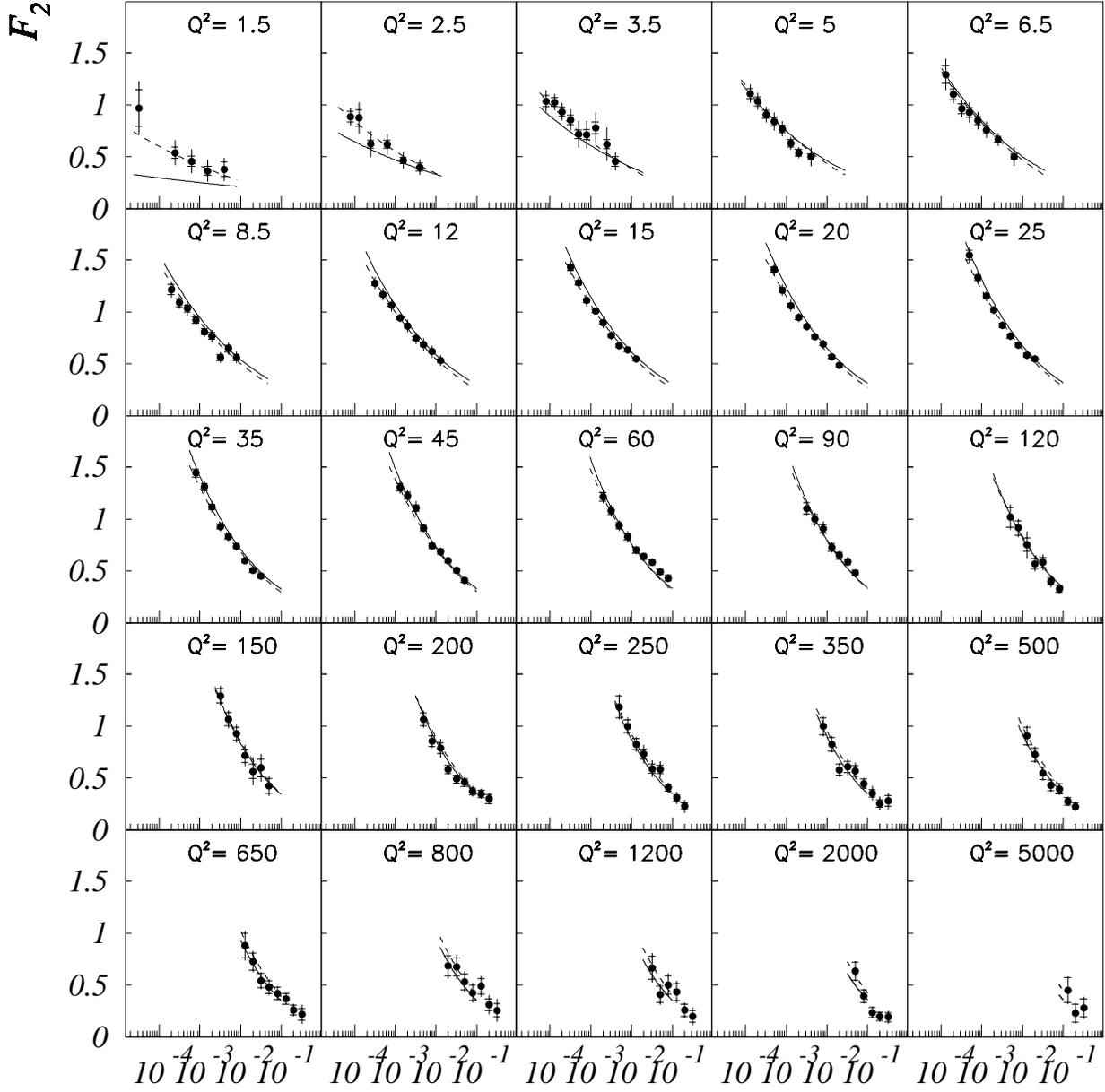,width=17cm,bbllx=0pt,bblly=0pt,bburx=540pt,bbury=802pt}}
\end{picture}
\end{center}
\caption[]{\sl The proton structure function \F as measured by the H1
experiment at HERA together with a fit to the NLO double asymptotic expression
(1) (full line) for $Q^2>5$ GeV$^2$ and 
with a fit  to the modified DLL expression 
$F_2 = N_fe^{-\gamma\sqrt{T/\xi}}$ (dashed line)
in the full $Q^2$ range. }
\end{figure}
\end{document}